\newcommand{\PRL}[3]{{\it{Phys.}}\ {\it{Rev.}}\ {\it{Lett.}}\ {\bf #1},\ #2 (#3)}
\newcommand{\RMP}[3]{{\it{Rev.}}\ {\it{Mod.}}\ {\it{Phys.}}\ {\bf #1},\ #2 (#3)}

\newcommand{\NAT}[3]{{\it{Nature}}\ {\bf #1},\ #2 (#3)}
\newcommand{\SC}[3]{{\it{Science}}\ {\bf #1},\ #2 (#3)}

\newcommand{\PRA}[3]{{\it{Phys.}}\ {\it{Rev.}}\ A\ {\bf #1},\ #2 (#3)}

\newcommand{\PRE}[3]{{\it{Phys.}}\ {\it{Rev.}}\ E\ {\bf #1},\ #2 (#3)}

\newcommand\ra{\rightarrow}

\newcommand{\diracslash}[1]{#1\llap{/\kern2pt}}

\newcommand{\be}{\begin{equation}}
\newcommand{\ee}{\end{equation}}
\newcommand{\bea}{\begin{eqnarray}}
\newcommand{\eea}{\end{eqnarray}}
\newcommand{\ba}[1]{\begin{array}{#1}}
\newcommand{\ea}{\end{array}}

\newcommand\sech{\mathrm{sech}}
\documentclass[12pt,a4paper]{article}
\usepackage[dvips]{graphicx}
\begin{document}
\begin{center}
\large{\bf Complex Envelope Soliton in Bose-Einstein Condensate with Time
Dependent Scattering Length}\\
\vspace{0.4in}
{\it Ayan Khan}\footnote{akhan@prl.res.in},
{\it Rajneesh Atre} \footnote{atre@prl.res.in} and
{\it Prasanta K. Panigrahi}\footnote{prasanta@prl.res.in}\\
Physical Research Laboratory, Navrangpura,\\
Ahmedabad, 380 009, India\\
\end{center}
{\small We elaborate on a general method to find complex envelope
solitons in a cigar shaped Bose-Einstein condensate in a trap. The
procedure incorporates time dependent scattering length, oscillator
frequency and loss/gain. A variety of time dependencies of the above
parameters, akin to the ones occurring in the experiments can be tackled. \vspace{0.2in}}\\

Coherent atom optics is a subject of much current interest for its
relevance to both basic physics and technology. For this purpose,
lower dimensional condensates {\em{e.g.}}, cigar shaped
Bose-Einstein condensates (BECs) have been the subject of active
study in recent years  \cite{dalafovo}. Intense investigations about
the behavior of the condensate in the presence of time varying
control parameters like, scattering length, oscillator frequency
{\em{etc.}}, are being carried out for the purpose of optimal
control. In BEC, solitary waves have been experimentally observed,
both in repulsive and attractive domains \cite{khaykovich,strecker}.
These have been understood from the non-linear Gross-Pitaevskii (GP)
equation describing weakly coupled BEC, \bea\label{eq:GP} i\hbar
\frac{\partial \Psi({\bf{r}},t)}{\partial
t}&=&\left\{-\frac{\hbar^2}{2m}\nabla^{2}+U|\Psi({\bf{r}},t)|^{2}+V\right\}\Psi({\bf{r}},t)\textrm{,}
\eea where $U=4\pi\hbar^2a_{s}(t)/m$, $a_{s}=$ the s-wave scattering
length,
and $V$ is the external potential.\\
The scattering length, which is the coefficient of nonlinearity in
GP equation, can be adjusted both in sign and magnitude through
Feshbach resonance. Recently a family of exact solutions of
quasi-one dimensional GP equation with time varying scattering
length, loss/gain, in the presence of oscillator has been obtained
\cite{atre}. These are chirped solitons with real envelope. In
recent years, solitons with complex envelope is generating interest
in different areas of physics. These are known as Bloch solitons in
condensed matter physics. In optical fibers one also finds these
solitons. In BEC, complex solitons have been investigated
\cite{jack,jackson,komi,pop,kom}. Very recently complex envelope
solitons in optical lattices has been reported \cite{porter}.

In the present study, we explicate a method to obtain general
complex envelope solitons in cigar shaped BEC, in a trap
incorporating time dependent scattering length, oscillator
frequency, and loss/gain. We have constructed the two hydrodynamic
equations in this scenario and have found their solutions.
Our analysis starts with quasi-one dimensional non-linear Schr\"odinger equation (NLSE)
derived from three dimensional GP equation with an
additional loss/gain term $g(t)$ and time dependent chemical potential
$\nu(t)$. In dimensionless units this is given by \cite{atre},
\begin{eqnarray}\label{eq:GP1}
i\partial_{t}\psi&=&-\frac{1}{2}\partial_{zz}\psi+\frac{1}{2}M(t)z^{2}\psi+
\gamma(t)|\psi|^{2}\psi+\frac{ig(t)}{2}\psi+\frac{1}{2}\nu(t)\psi \textrm{.}
\end{eqnarray}
We assume the following ansatz solution \bea\label{eq:ansatz}
\psi(z,t)&=&\sqrt{A(t)}F\left[A(t){z-l(t)}\right]\exp\left[{i\{\chi(z,t)+\phi(z,t)\}
+\frac{G(t)}{2}}\right]\textrm{,}
\eea where $\sqrt{A(t)}F[A(t){z-l(t)}]\exp[i\chi(z,t)]$ carries the
signature of complex envelope. Further $G(t)$ and $l(t)$ have been defined
as, \bea\label{eq:loss} G(t)&=&\int_{0}^{t}{g(t')dt'}\textrm{ ,
}\,\,l(t)=\int_{0}^{t}{v(t')dt'}\textrm{.} \eea The phase is chirped as,
\begin{eqnarray}\label{eq:chirp}
\phi(z,t)&=&a(t)+b(t)z-\frac{1}{2}c(t)z^{2}\,\,\textrm{, where},~~
a(t)=a_{0}-\frac{1}{2}\int_{0}^{t}A^{2}(t')dt'\textrm{.} \nonumber
\end{eqnarray}
Substitution of the above ansatz in Eq.(\ref{eq:GP1}) allows one
to separate this equation in imaginary and real parts.
The knowledge of the earlier conditions, leads one to write the
imaginary equation as,
\bea\label{eq:im1}
-uF'&=&-F'\chi'-\frac{1}{2}F\chi'' \textrm{, and}\nonumber\\
\chi'&=&u-\frac{2C_{0}}{F^{2}}\textrm{.} \eea Here prime denotes
derivative with respect to $T$, where $T=A(t)[z-l(t)]$. We have set
$\frac{l_{t}+cl-A}{A}=u$, $A_{t}=Ac$ and $C_{0}$ is an integration
constant. Redefining $C_{0}=\frac{uF_{0}^{2}}{2}$, $F=\sqrt{\sigma}$ and
$F_{0}=\sqrt{\sigma_{0}}$, Eq.(\ref{eq:im1}) leads to, \be\label{eq:hyd1}
\chi'=u\left(1-\frac{\sigma_{0}}{\sigma}\right)\textrm{.} \ee We have
imposed the boundary condition, $\chi'\ra 0$ for $\sigma\ra\sigma_{0}$.
The local phase velocity of the solitary pulse is described by $\chi'$ and
local density is described by $\sigma$, $\sigma_{0}$ represents
equilibrium
density.\\
From the real equation one can extract a Riccati type equation,
$c_{t}-c^{2}(t)=M(t)$ which yields the solution of $c(t)$. The real
equation also yields, \bea\label{eq:real1} F''-\mu F-2\kappa
F^{3}&=&\chi'^{2}F-2u\chi'F \textrm{,} \eea where
$\kappa=\frac{\gamma_{0}}{A_{0}}$. Applying the following
consistency conditions, one obtains \bea
\gamma(t)&=&\gamma_{0}e^{-G}A(t)/A_{0}\textrm{ ,}\,\, b(t)=A(t)\textrm{,}\nonumber\\
A(t)&=&A_{0}\exp\left[{\int_{0}^{t}c(t')dt'}\right]\textrm{,}\,\,\nu(t)=A^{2}\frac{\mu}{2}
\textrm{,}\nonumber \eea and, \be\label{eq:real2} F''-\epsilon
F-2\kappa F^{3}-\frac{\lambda}{F^{3}}=0\textrm{,} \ee where
$\epsilon=\mu-u^{2}$ and $\lambda=u^{2}F_{0}^{4}$. Integration of
Eq.(\ref{eq:real2}) leads to the convenient form,
\begin{eqnarray}\label{eq:hydro2}
\Bigg(\frac{\partial \sqrt{\sigma}}{\partial T}\Bigg)^{2}&=&(\kappa \sigma -u^{2})
\frac{(\sigma-\sigma_{0})^{2}}{\sigma} \textrm{.}
\end{eqnarray}
The solution for Eq.(\ref{eq:hydro2}) is,
\begin{eqnarray}\label{eq:solution}
\sigma(z,t)=\sigma_{0}\left\{1-\cos^{2}{\theta}~
\sech^{2}\,{\left[\frac{A(z-l)\cos{\theta}}{\zeta}\right]}\right\}
\textrm{.}
\end{eqnarray}
These are the dark and grey solitons in the distributed scenario.
\begin{figure}[!h]
\centering
\includegraphics[height=5.2cm,width=5.2cm]{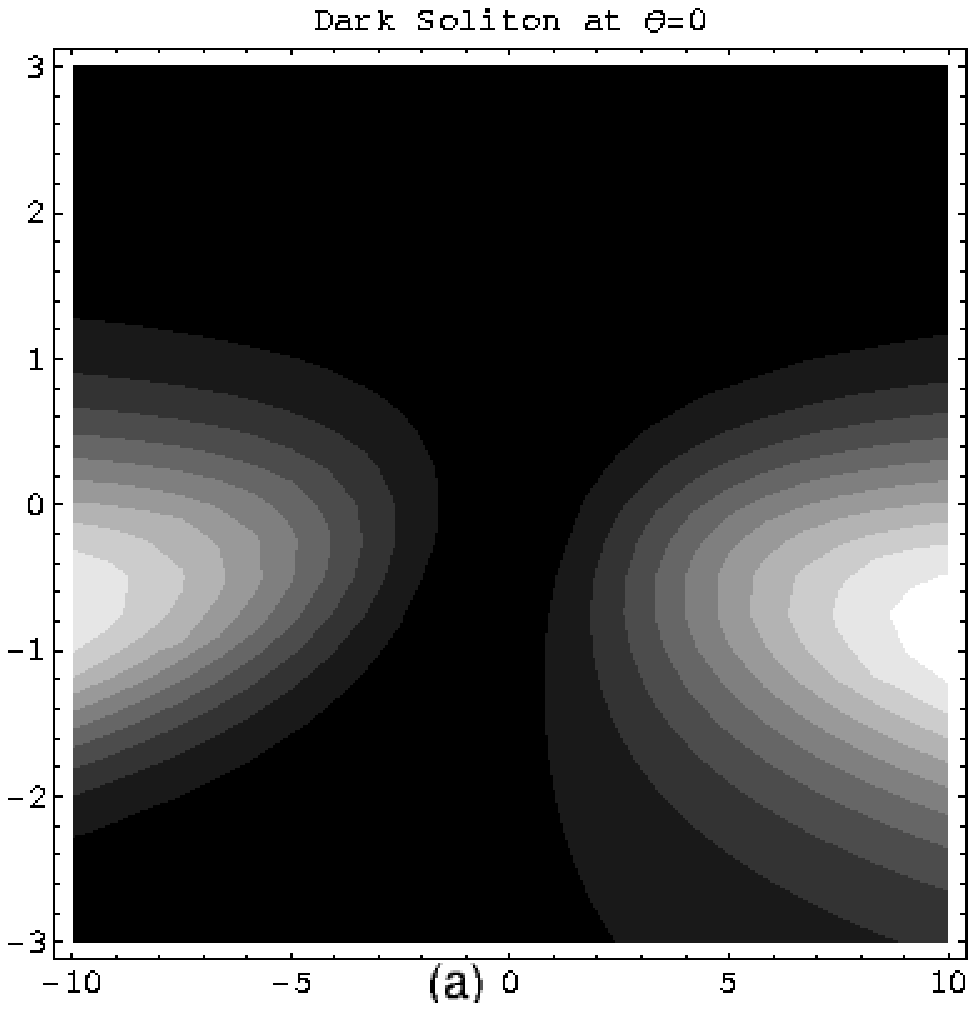}
\hspace{0.9in}
\includegraphics[height=5.7cm,width=5.7cm]{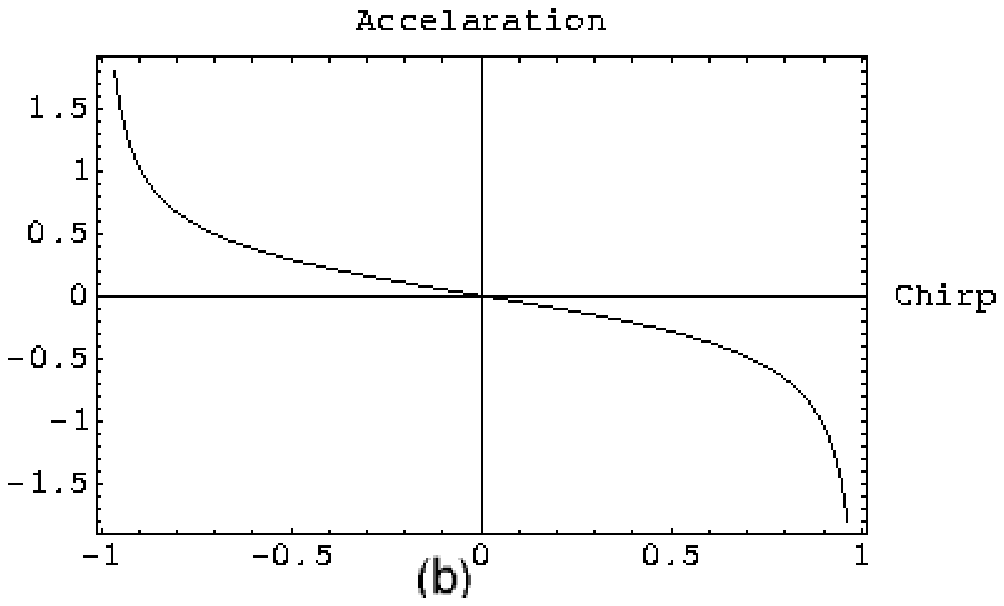}
\caption{(a) Density profile of the Condensate and (b) Chirp Controlled Soliton Motion}
\end{figure}

In conclusion, for quasi-one dimensional GP equation in an
oscillator potential with time dependent coupling and loss/gain, a
general complex dark soliton profile is obtained. In the presence of
expulsive oscillator the motion of the condensate is controlled by
the sign of the chirp, {\em{i.e}}, whether the soliton is
accelerating or decelerating.

\end{document}